\documentclass[twoside,twocolumn,english]{revtex4}
\usepackage[T1]{fontenc}
\usepackage[latin1]{inputenc}
\usepackage{geometry}
\usepackage{babel}
\usepackage{graphics}

\makeatletter

\providecommand{\LyX}{L\kern-.1667em\lower.25em\hbox{Y}\kern-.125emX\@}

\makeatother
\begin{document}

\title{{\LARGE Magnitude distribution of earthquakes: Two fractal contact
area distribution}}

\author{Srutarshi Pradhan\( ^{a} \)}

\email{spradhan@cmp.saha.ernet.in}

\author{Bikas K. Chakrabarti\( ^{a} \)}

\email{bikas@cmp.saha.ernet.in}

\author{Purussatam Ray\( ^{b} \)}

\email{ray@imsc.res.in}

\author{Malay Kanti Dey\( ^{c} \)}

\affiliation{\( ^{a} \)Condensed Matter Physics Group, Saha Institute of Nuclear
Physics, 1/AF , Bidhan Nagar, Kolkata -700 064, India}

\affiliation{\( ^{b} \)Institute of Mathematical Sciences, Chennai, India}

\affiliation{\( ^{c} \)Variable Energy Cyclotron Center, 1/AF , Bidhan Nagar,
Kolkata -700 064, India}

\begin{abstract}
\noindent The `plate tectonics' is an observed fact and most models
of earthquake incorporate that through the frictional dynamics (stick-slip)
of two surfaces where one surface moves over the other. These models
are more or less successful to reproduce the well known Gutenberg-Richter
type power law in the (released) energy distribution of earthquakes.
During sticking period, the elastic energy gets stored at the contact
area of the surfaces and is released when a slip occurs. Therefore,
the extent of the contact area between two surfaces plays an important
role in the earthquake dynamics and the power law in energy distribution
might imply a similar law for  the contact area distribution. Since,
fractured surfaces are fractals and tectonic plate-earth's crust interface
can be considered to have fractal nature, we study here the contact
area distribution between two fractal surfaces. We consider the overlap
set (\( m \)) of two self-similar fractals, characterised by the
same fractal dimensions (\( d_{f} \)), and look for their distribution
\( P(m) \). We have studied numerically the specific cases of both
regular and random Cantor sets (in the embedding dimension \( d=1 \)),
gaskets and percolation fractals (in \( d=2 \)). We find that in
all the cases the distributions show an universal finite size (\( L \))
scaling behavior \( P^{\prime }(m^{\prime })=L^{\alpha }P(m,L) \);
\( m^{\prime }=mL^{-\alpha } \), where \( \alpha =2(d_{f}-d) \).
The \( P(m) \), and consequently the scaled distribution \( P^{\prime }(m^{\prime }) \),
have got a power law decay with \( m \) (with decay exponent equal
to \( d \)) for both regular and random Cantor sets and also for
gaskets. For percolation clusters, \( P(m) \) (and hence \( P^{\prime }(m^{\prime }) \))
have a Gaussian variation with \( m \). 
\end{abstract}
\maketitle

\section{\textbf{Introduction}}

\noindent \vskip -.2in

\noindent Enormous efforts have been made by geologists and physicists
to understand the earthquake phenomena since several decades. Its
dynamics is still a challenging problem. Though model studies of earthquake
have been going on several decades now, there is no consensus regarding
one single model. `Plate-tectonics' is an important observation in
this context by the geologists and the Gutenberg-Richter power law
is an unique characterisation of this dynamics. The earth's solid
outer crust (about 20 km thick) rests on a tectonic shell. This tectonic
shell is divided into a number (about 12) of mobile plates, having
relative velocities of the order of few centimeters per year. This
motion of the plates arises due to the powerful convective flow of
the earth's mantle, at the inner core of earth. Due to the surface
roughness of both the earth's crust and the plates, solid-solid frictional
forces arise and this helps sticking the earth's crust to the plates.
This sticking develops elastic strains and the strain energy gradually
increases because of the uniform motion of the tectonic plates. There
is a competition between the sticking frictional force and the restoring
elastic force (stress). When the accumulated stress exceeds the frictional
force, a slip (earth quake) occurs and it releases the stored additional
elastic energy. Gutenberg and Richter (1954) \cite{GR54}, by analysing
the records, observed a power-law distribution of the elastic energy
released during earthquakes given as \( N(\epsilon )\sim \epsilon ^{-\alpha } \),
where \( N(\epsilon ) \) is the number of earthquakes releasing energy
greater than or equal to \( \epsilon  \) and \( \alpha  \) is the
power exponent. The observed value of \( \alpha  \) ranges between
0.7 and 1.0. Several hypothesis and model systems have been proposed
to investigate the nature of the earthquake phenomena \cite{CB97}.
The main intention of the model studies is to capture the above Gutenberg-Richter
type power law for the frequency distribution of failures (quakes)
in the failure dynamics of the models. Among these, most of the models
\cite{BK67,CL89} incorporate the stick-slip process and the roughness
\cite{Hansen03} of the surfaces involved as important features. Recently
some models \cite{V96,BS99} have included the fractal nature of both
the earth's crust and the plate involved in stick-slip process and
initiated new modellings of earthquake dynamics. As the elastic energy
can be stored at the contact area of the surfaces only, the contact
area distribution should have much importance. Also, in fractal physics
the statistics of two fractal overlap is still missing although this
may be useful to study the interface properties in many physical situations. 

In this report we have studied the contact area distributions and
their scaling properties for different sets of fractal surfaces, using
computer simulation techniques.

\newpage

\section{Simulation studies of two fractal contact area distribution}

\subsection{\noindent Model}

\noindent Extensive studies have already established different types
of fractals and their properties \cite{BS95,Bak97,Man82}. The statistics
of overlaps between two such fractals is however not studied much
yet, though their knowledge is often required in various physical
contexts. For example, it has been claimed that since the fractured
surfaces have got well-characterized self-affine properties, the distribution
of the elastic energies released during the slips between two fractal
surfaces (earthquake events) may follow the overlap distribution of
two self-similar fractal surfaces \cite{V96,BS99,BMS00}. Also, using
renormalisation group technique Chakrabarti and Stinchcombe \cite{BS99}
have analytically shown that for regular fractal overlap (Cantor sets
and carpets) the contact area distribution follows power law. 

Here, we study the distribution \( P(m) \) of contact area \( m \)
between two well-characterized fractals having the same fractal dimension.
We have chosen different types of fractals: regular or non-random
Cantor sets, random Cantor sets (in one dimension), regular and random
gaskets on square lattice and percolating clusters \cite{Stauffer 92,Leath 76}
embedded in two dimensions. We find a universal scaling behavior of
the distribution: \begin{equation}
\label{may3}
P^{\prime }(m^{\prime })=L^{\alpha }P(m,L);m^{\prime }=mL^{-\alpha },
\end{equation}

\noindent where \( L \) denotes the finite size of the fractal and
the exponent \( \alpha =2(d_{f}-d) \); \( d_{f} \) being the mass
dimension of the fractal and \( d \) is the embedding dimension.
Also the overlap distribution \( P(m) \), and hence the scaled distribution
\( P^{\prime }(m^{\prime }) \), seen to decay with \( m \) or \( m^{\prime } \)
following a power law (with exponent value equal to the embedding
dimension of the fractals) for both regular and random Cantor sets
and gaskets: \begin{equation}
\label{sep25}
P(m)\sim m^{-\beta };\beta =d.
\end{equation}
For the percolating clusters, however, the overlap distribution takes
a Gaussian form. It may be noted that the normalisation (to unity)
restriction on both \( P \) and \( m \) ensures the same scaling
exponent \( \alpha  \) for both.

\subsection{\noindent Overlaps between regular fractals}

\noindent Here we construct three types of regular fractals: regular
Cantor sets of dimension \( \ln 2/\ln 3 \), regular Cantor sets of
dimension \( \ln 4/\ln 5 \) and regular gaskets of dimension \( \ln 3/\ln 2 \)
on a square lattice. These regular fractals are constructed following
a repetitive procedure in successive generations (\( n \)) such that
the self-similarity is strictly maintained at every stage. For example,
a regular Cantor set of dimension \( \ln 2/\ln 3 \) is formed in
the \( n\rightarrow \infty  \) limit of the set obtained by removing
the middle one-third portion of each occupied set at every generation,
starting from a compact set of size \( L=3^{n} \) at \( n=1 \) (see
Fig. 1). For the set with dimension \( \ln 4/\ln 5 \), one similarly
removes the middle one-fifth portion at each generation (here \( L=5^{n} \)
for finite \( n \)-th generation). 

\resizebox*{7cm}{3cm}{\includegraphics{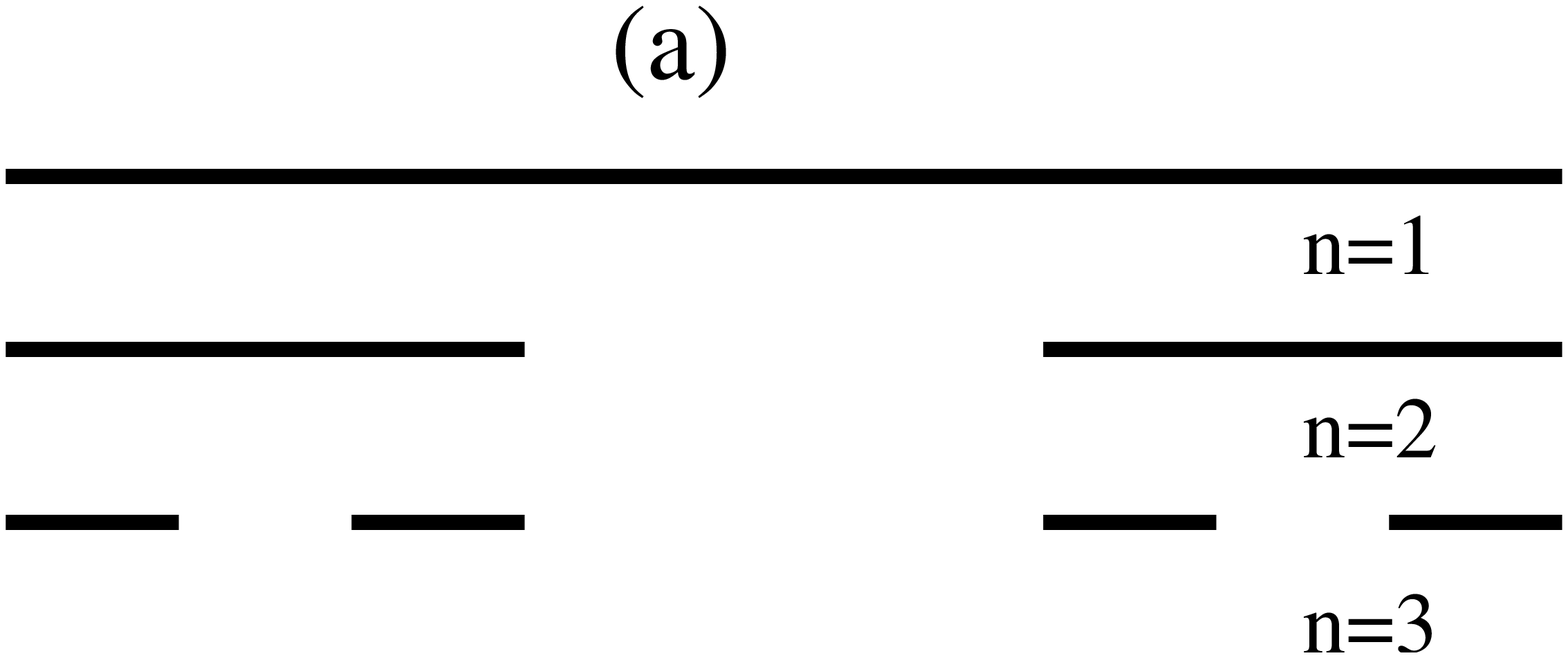}}  

\resizebox*{7cm}{3cm}{\includegraphics{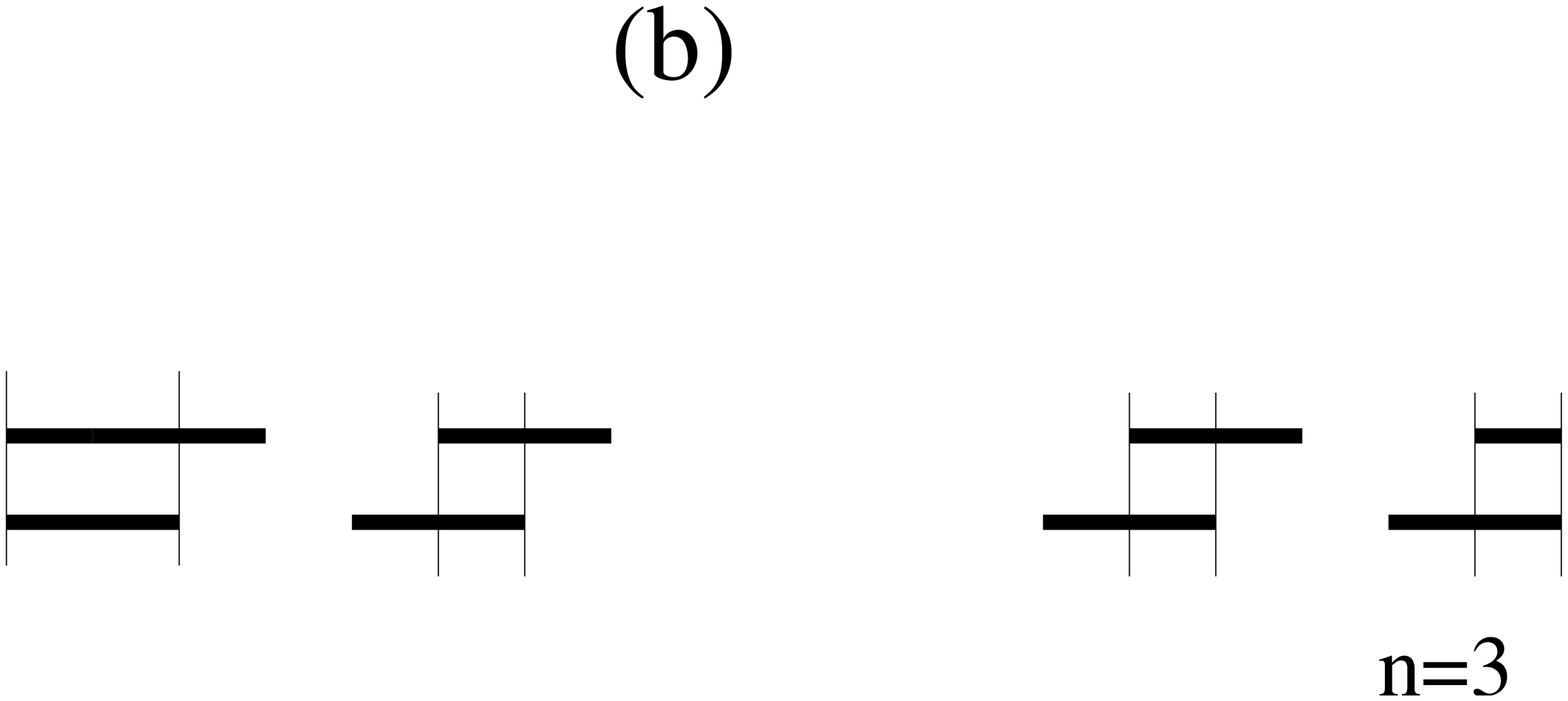}} 

\resizebox*{6cm}{5.5cm}{\rotatebox{-90}{\includegraphics{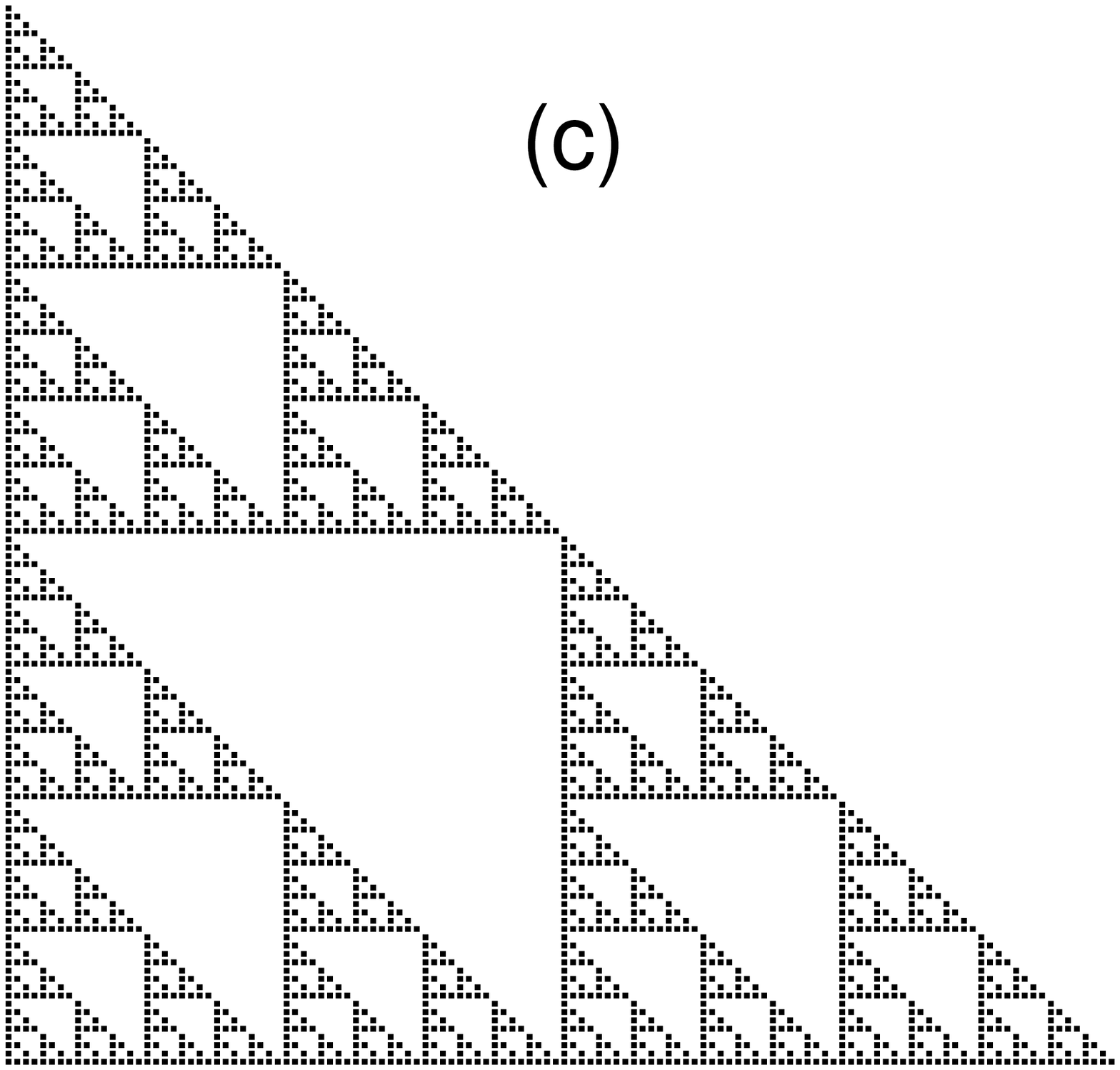}}}  

\resizebox*{6cm}{5.5cm}{\rotatebox{-90}{\includegraphics{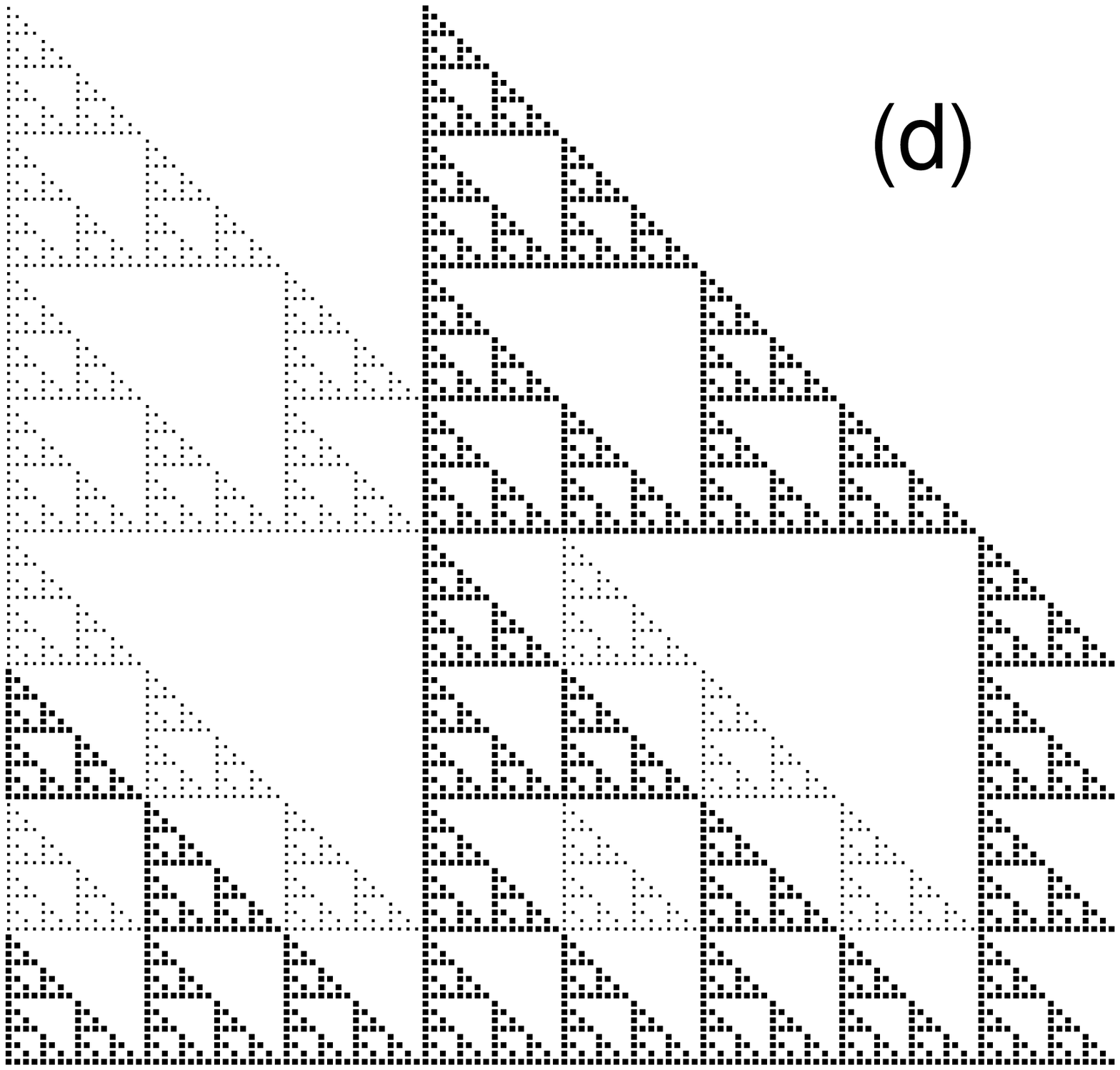}}} 

\vskip.1in

\noindent {\footnotesize Fig 1}\textbf{\footnotesize :} {\footnotesize (a)
A regular Cantor set of dimension \( \ln 2/\ln 3 \); only three finite
generations are shown. (b) The overlap of two identical (regular)
Cantor sets, at \( n=3 \), when one slips over other; the overlap
sets are indicated within the vertical lines, where periodic boundary
condition has been used. (c) A regular gasket of dimension \( \ln 3/\ln 2 \)
at the \( 7 \)th generation. (d) The overlap of two identical regular
gaskets at same generations (\( n=7 \)) is shown as one is translated
over the other; periodic boundary condition has been used for the
translated gasket. }{\footnotesize \par}

To study the overlap between two such identical fractals, the boundary
effects are avoided using periodic boundary condition for one and
we consider the other set to slip over the first. At each step of
such slip or translation, we count the overlapping filled sites (black
parts or dots) present in both the fractals and the total number of
such sites gives the size of overlap \( m \). Thus, if one sequentially
translates one fractal over another, various overlap values (\( m \))
are obtained which in turn give the distribution \( P(m) \). Since
no randomness is involved here, we do not need any configurational
averaging. These results are shown in Fig. 2, where the Cantor sets
and gaskets are generated for finite generations \( n \): \( L=3^{n} \)
for Cantor sets with \( d_{f}=\ln 2/\ln 3 \), \( L=5^{n} \) for
Cantor sets with \( d_{f}=\ln 4/\ln 5 \) and \( L=2^{n} \) for gaskets
with \( d_{f}=\ln 3/\ln 2 \). The overlap distributions \( P(m,L) \)
are fitted to the scaling forms (1) and (2). The results indicate
their validity in the large \( n \) (or \( L \)) limit.

\resizebox*{7cm}{6.5cm}{\rotatebox{-90}{\includegraphics{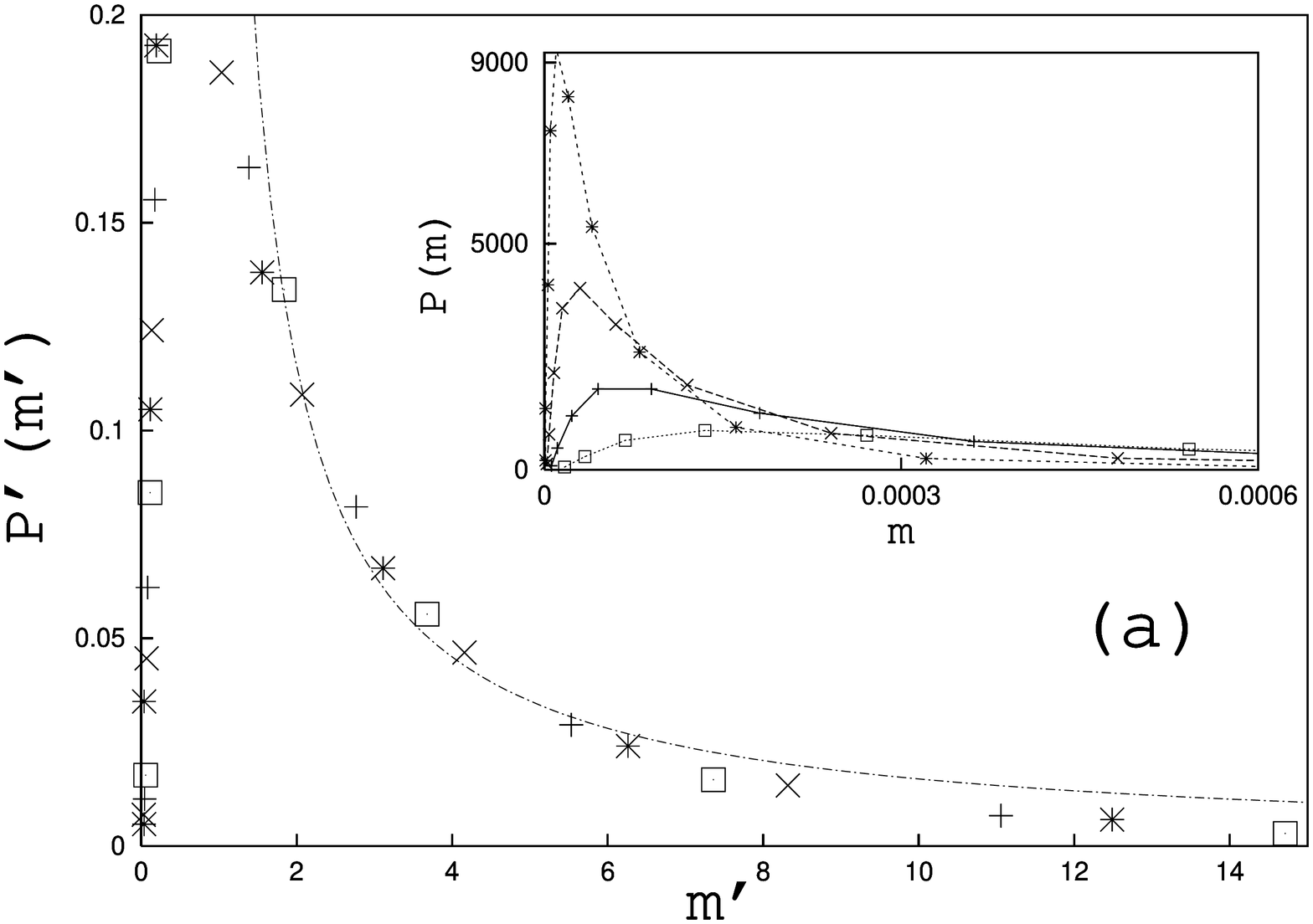}}} 

\resizebox*{7cm}{6.5cm}{\rotatebox{-90}{\includegraphics{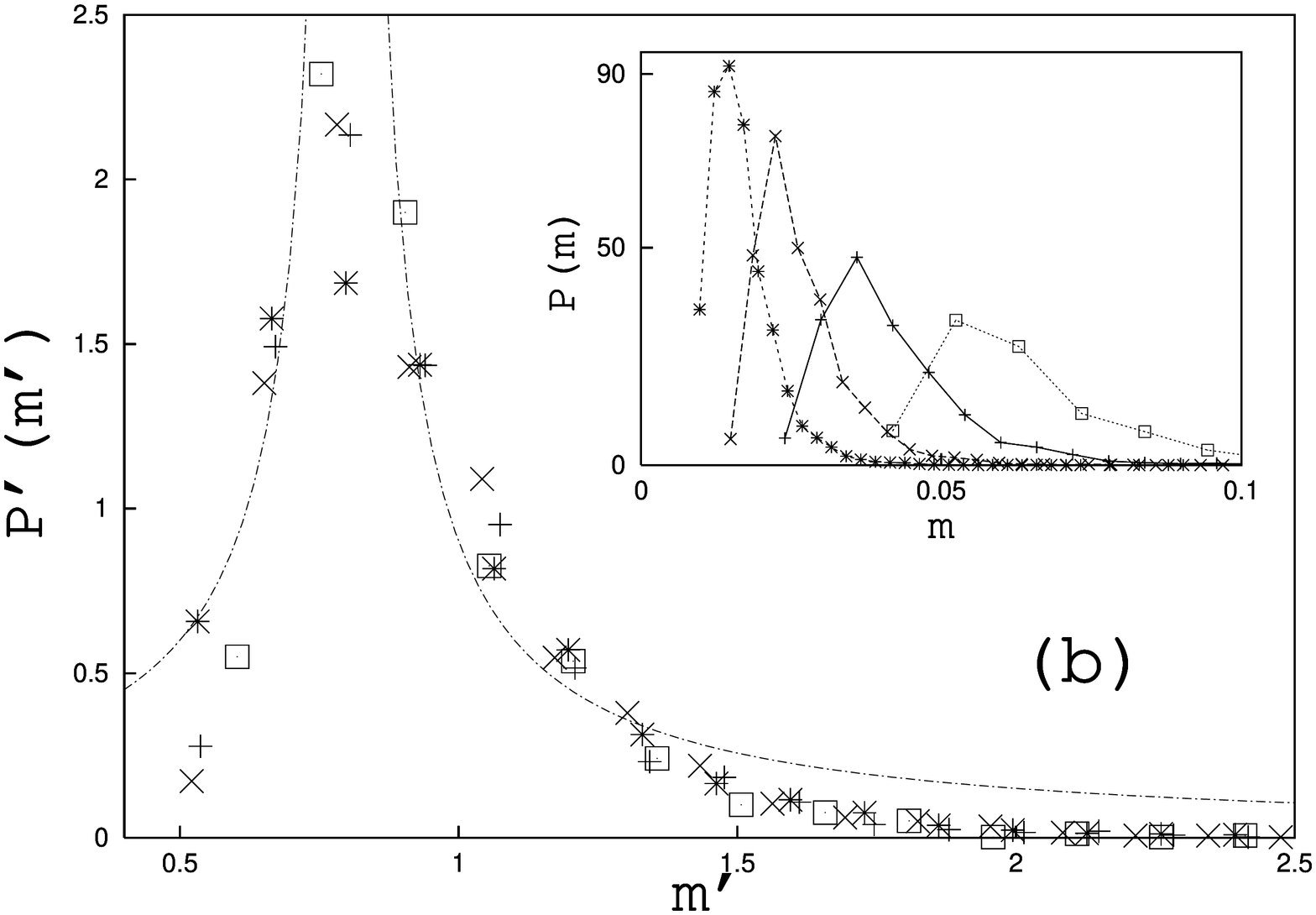}}} 

\vspace{0.3cm}
{\centering \resizebox*{7cm}{6.5cm}{\rotatebox{-90}{\includegraphics{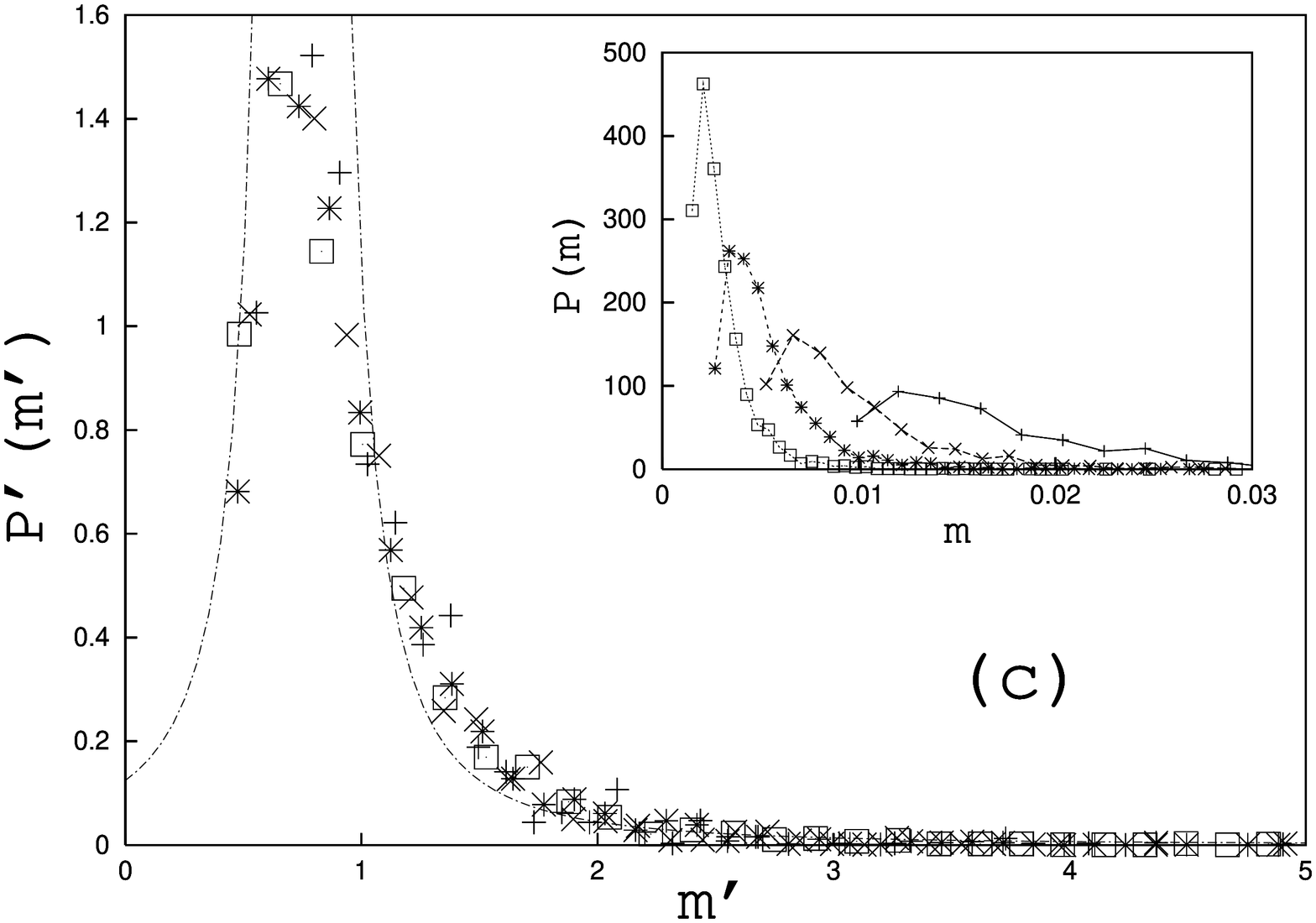}}} \par}
\vspace{0.3cm}

\noindent {\footnotesize Fig 2: The scaled distribution plot of \( P^{\prime }(m^{\prime })=P(m,L)L^{\alpha } \)
against the scaled overlap \( m^{\prime }=mL^{-\alpha } \) for two
identical regular fractals: (a) Cantor sets with \( d_{f}=\ln 2/\ln 3 \)
at various finite generations: \( n=10 \) (square), \( n=11 \) (plus),
\( n=12 \) (cross) and \( n=13 \) (star); (b) Cantor sets with \( d_{f}=\ln 4/\ln 5 \)
for finite generations: \( n=6 \) (square), \( n=7 \) (plus), \( n=8 \)
(cross) and \( n=9 \) (star); (c) gaskets with \( d_{f}=\ln 3/\ln 2 \)
for finite generations: \( n=7 \) (plus), \( n=8 \) (cross), \( n=9 \)
(star) and \( n=10 \) (square). Note that in all the three cases
\( \alpha =2(d_{f}-d) \) and the dotted lines indicate the best fit
curves of the form \( a(x-b)^{-d} \); where \( d \) is the embedding
dimension {[}\( d=1 \) for (a) and (b) and \( d=2 \) for (c){]}.
Insets show the unscaled distributions \( P(m) \) for overlap \( m \)
in different cases. }{\footnotesize \par}

\subsection{\noindent Overlaps between random fractals}

\noindent Here we construct three types of random fractals: random
Cantor sets of dimension \( \ln 2/\ln 3 \), random Cantor sets of
dimension \( \ln 4/\ln 5 \) and random gaskets of dimension \( \ln 3/\ln 2 \)
on square lattice. We can construct a random Cantor set of dimension
\( \ln 2/\ln 3 \) by removing randomly any of the one-third portion
at every stage or generation (see Fig 3).

\vskip .3in

\resizebox*{7cm}{3.5cm}{\includegraphics{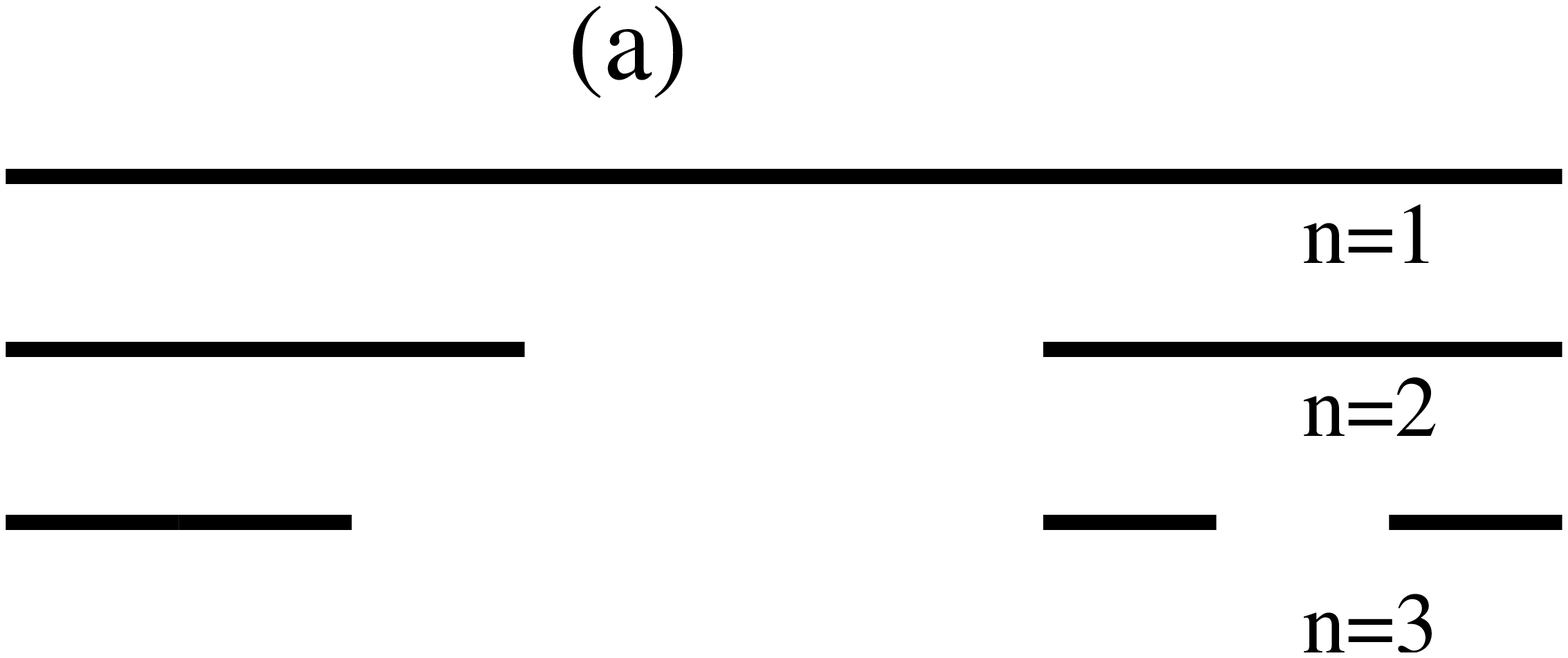}} 

\resizebox*{7cm}{3.5cm}{\includegraphics{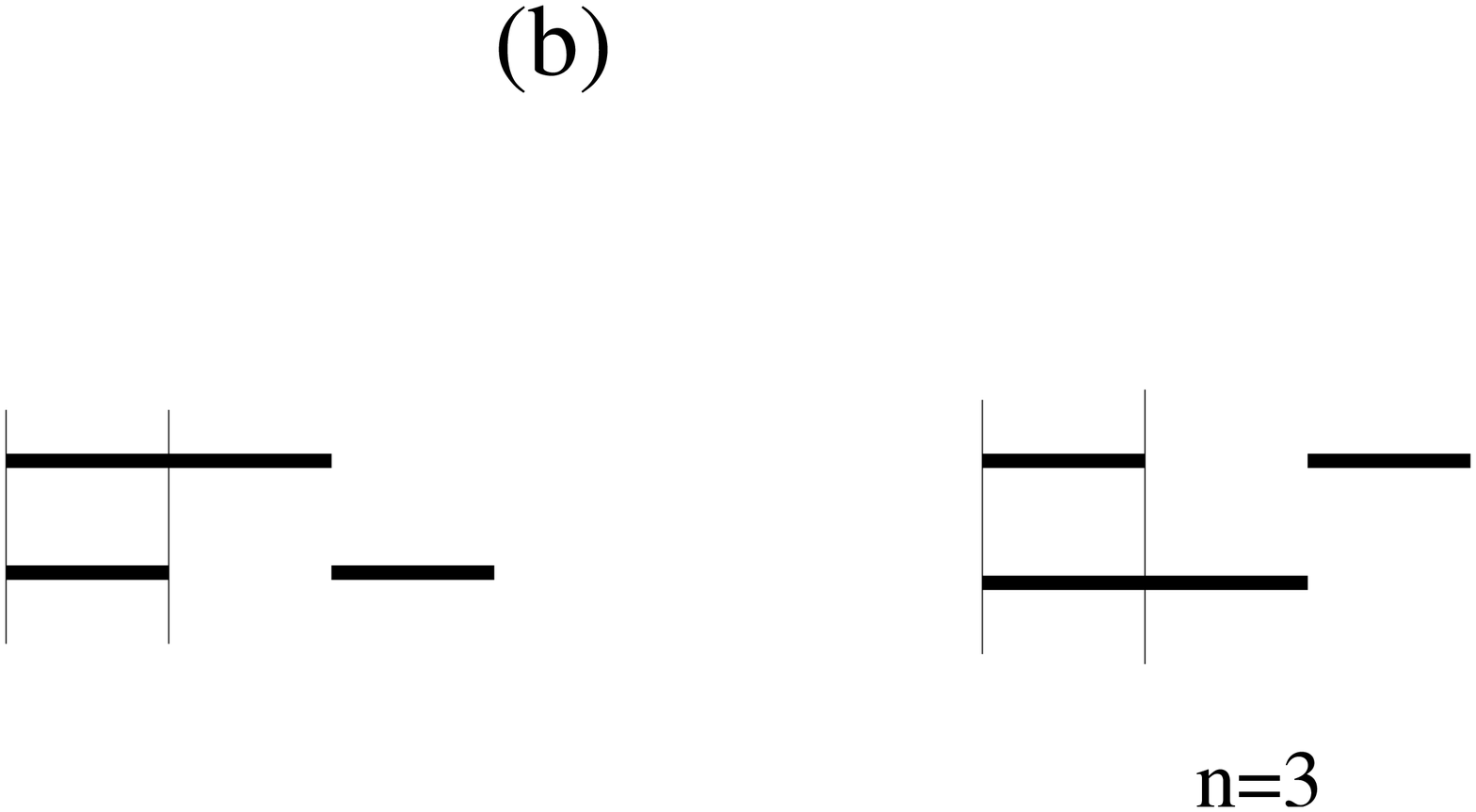}} 

\vskip.1in

\resizebox*{6cm}{6cm}{\rotatebox{-90}{\includegraphics{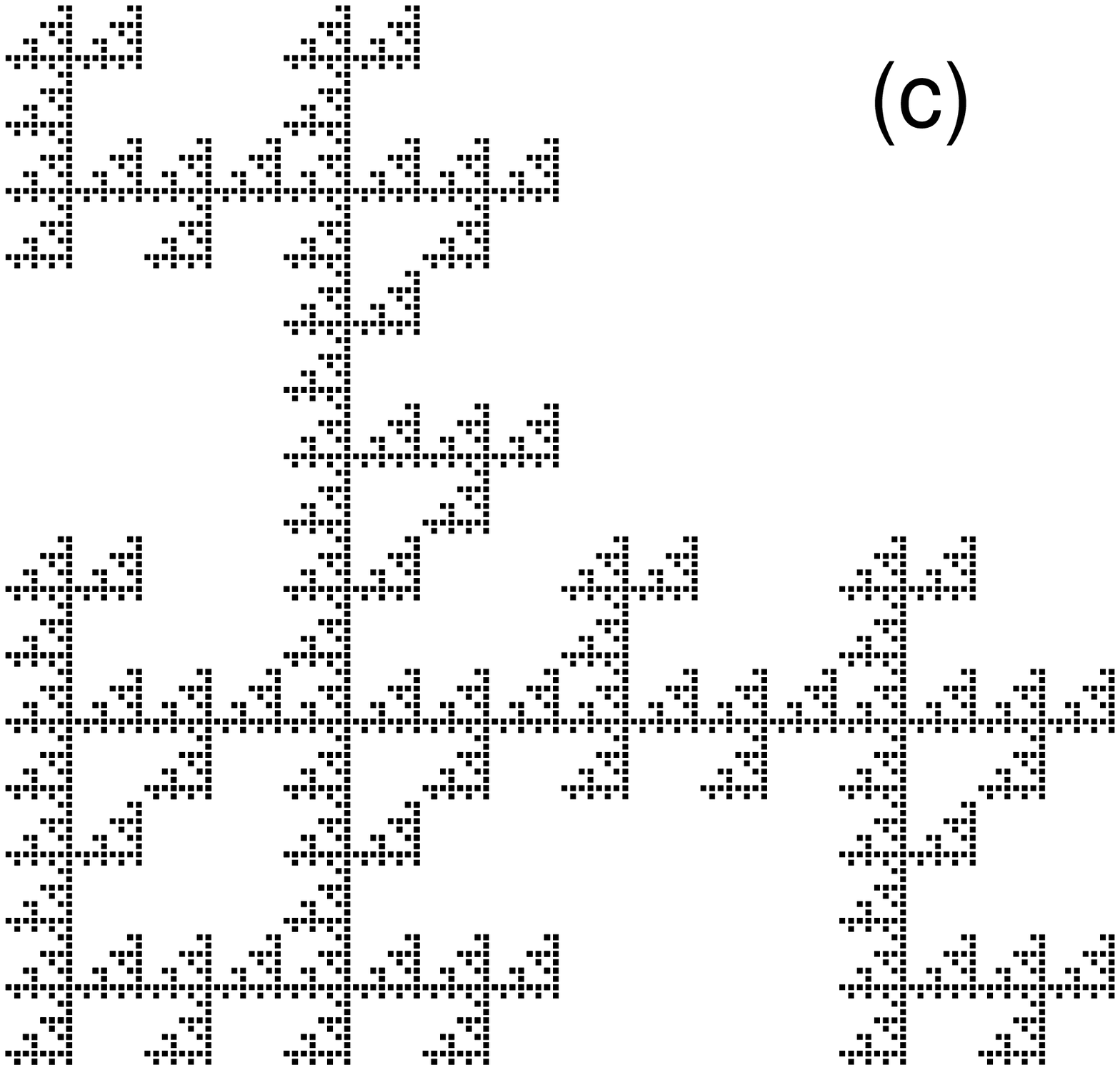}}} 

\resizebox*{6cm}{6cm}{\rotatebox{-90}{\includegraphics{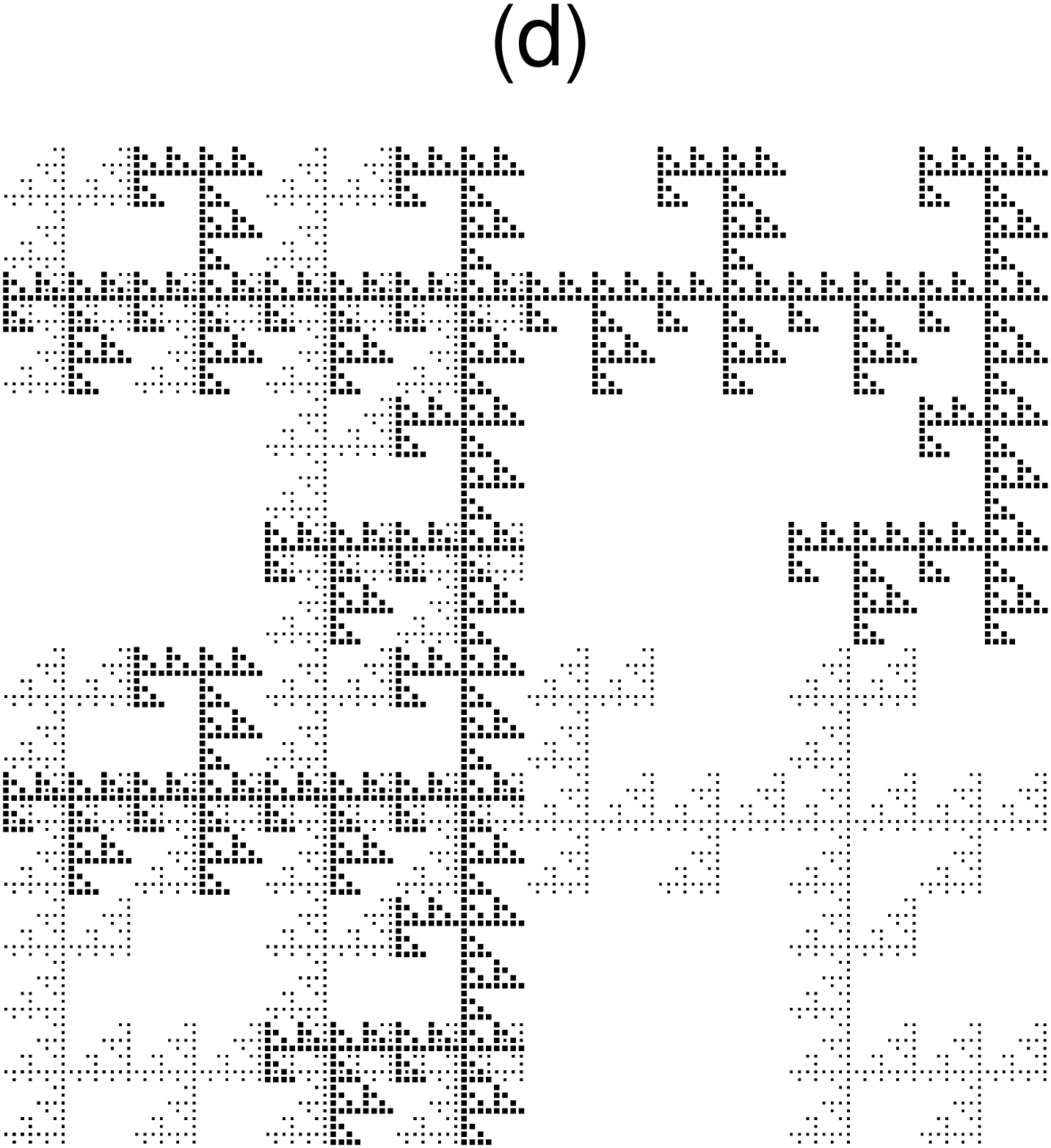}}} 

\noindent {\footnotesize Fig 3: (a) A random Cantor set of dimension
\( \ln 2/\ln 3 \); only three finite generations are shown. (b) Overlap
of two random Cantor sets (at \( n=3 \); having the same fractal
dimension) in two different realisations. The overlap sets are indicated
within the vertical bars. (c) A random realisation of a gasket of
dimension \( \ln 3/\ln 2 \) at \( 7 \)th generation. (d) The overlap
of two random gaskets of same dimension and of same generation but
generated in different realisations.}{\footnotesize \par}

Here the structure of the sets change with configurations as randomness
is involved. We therefore take the overlap between any two such sets
at finite generation \( n \) having same dimension but of different
configurations. We count the portions present in both the sets, and
this gives the total size of overlap \( m \). Clearly the overlap
sizes (\( m \)) change from one pair of configurations to other and
we determine their average distribution \( P(m) \). The finite size
\( L \) of the fractals is similarly related to generation number
\( n \) as discussed in the previous section. The average distributions
\( P(m,L) \) for finite size \( L \) of the fractals are determined
using \( 500 \) such configurations for each of the three types of
random fractals. 

\resizebox*{7cm}{6.2cm}{\rotatebox{-90}{\includegraphics{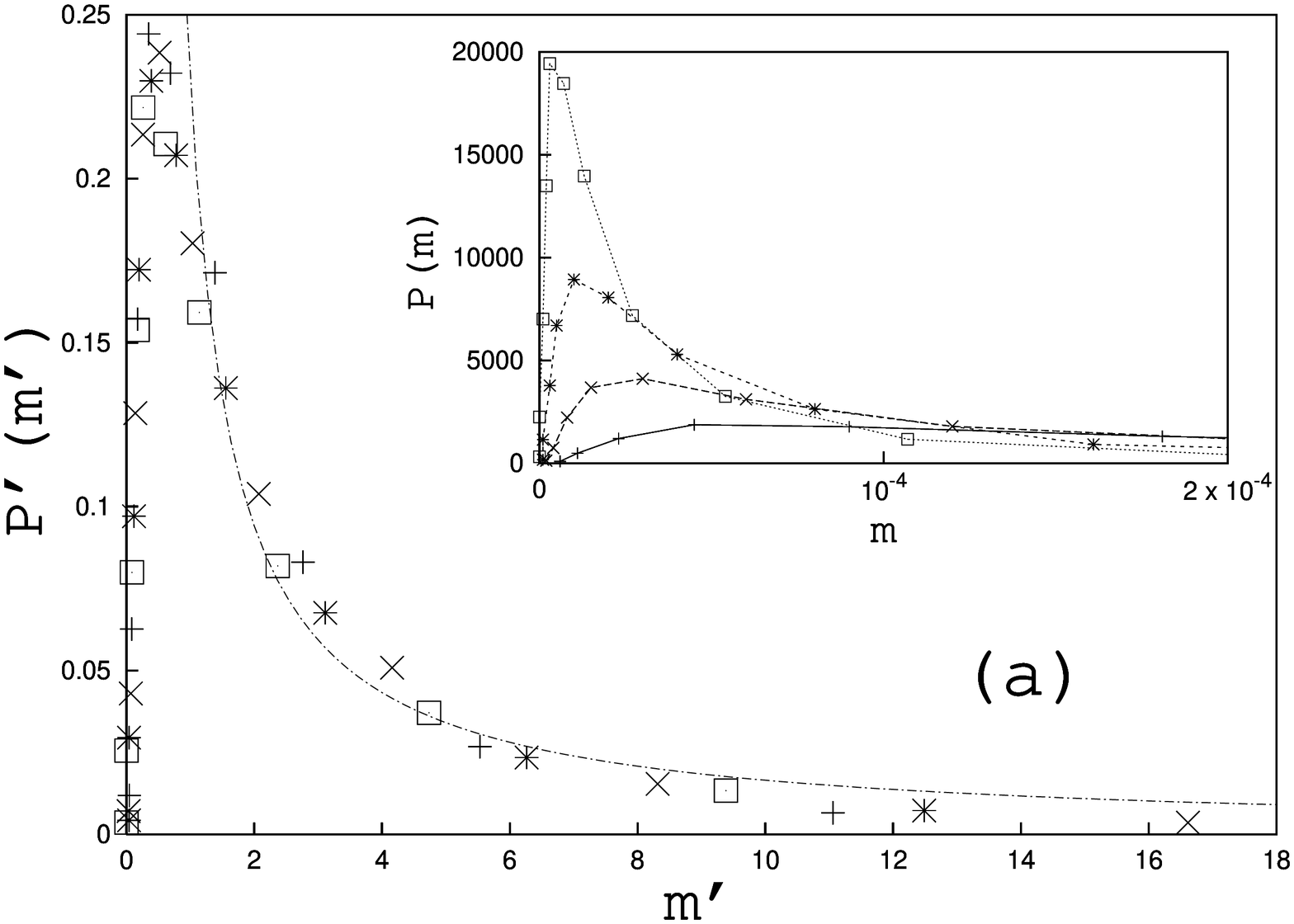}}} 

\resizebox*{7cm}{6.2cm}{\rotatebox{-90}{\includegraphics{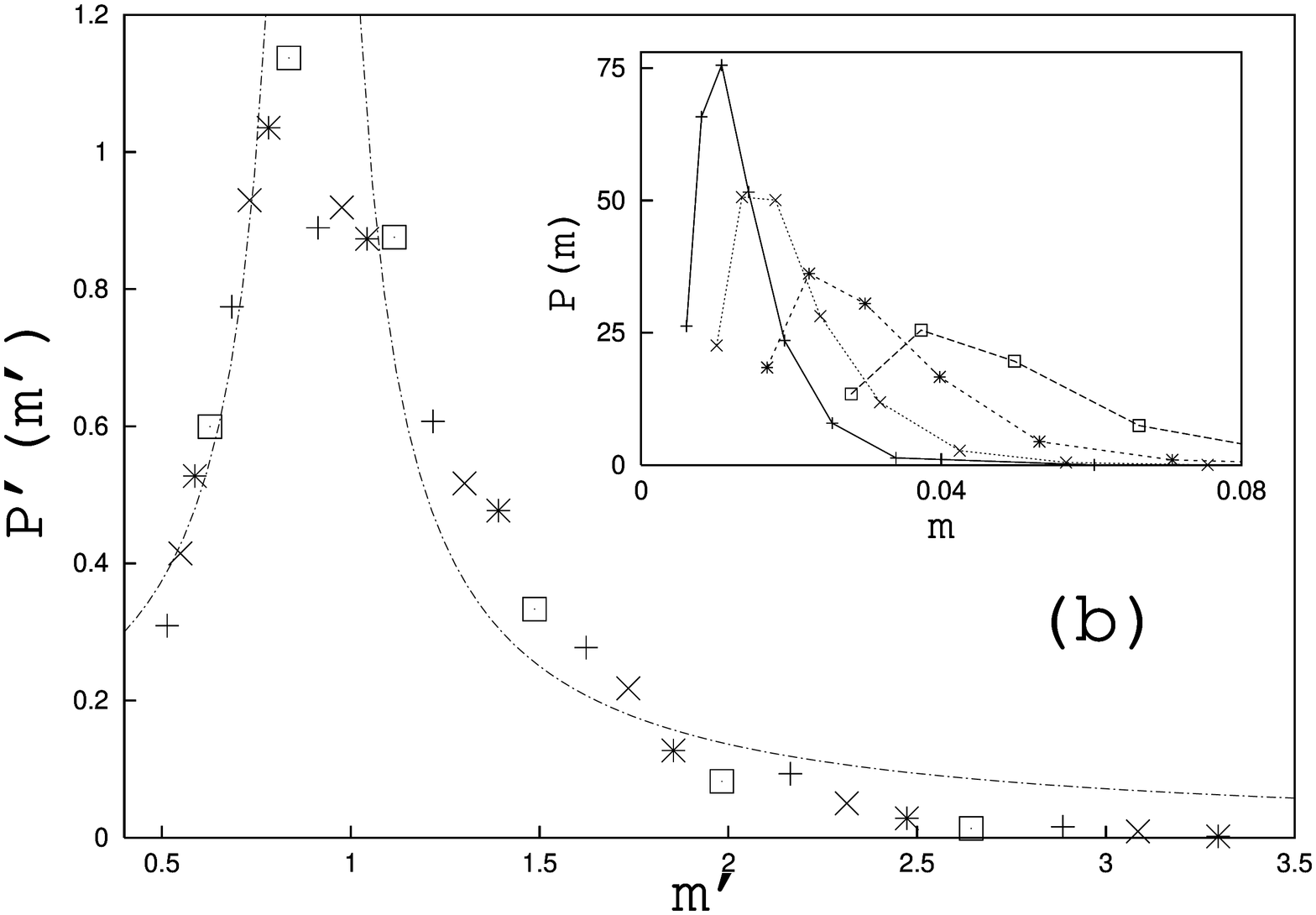}}} 

\vspace{0.3cm}
{\centering \resizebox*{7cm}{6.2cm}{\rotatebox{-90}{\includegraphics{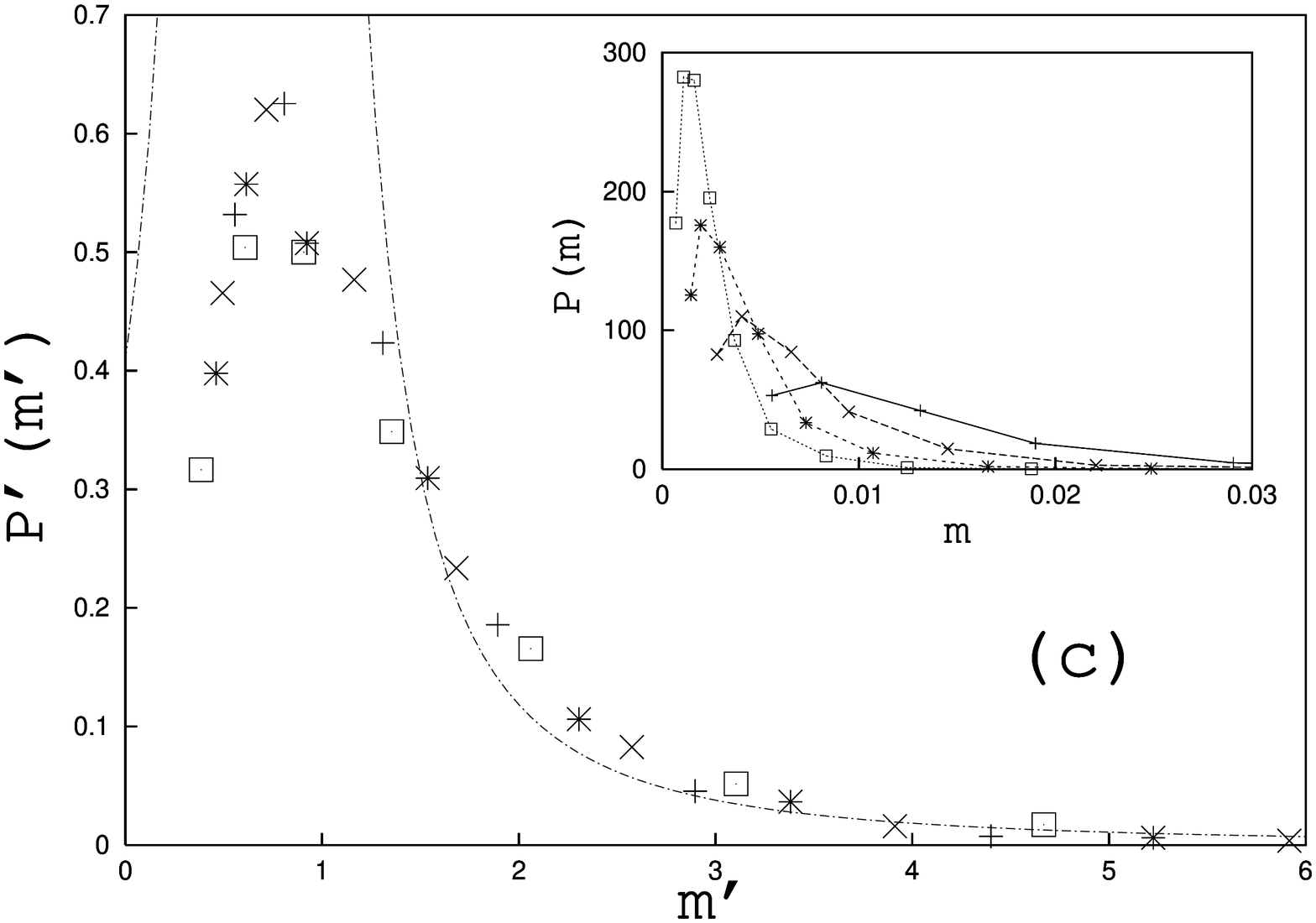}}} \par}
\vspace{0.3cm}

\noindent {\footnotesize Fig 4: The scaled distribution plot of \( P^{\prime }(m^{\prime })=P(m,L)L^{\alpha } \)
against the scaled overlap \( m^{\prime }=mL^{-\alpha } \) for random
fractals: (a) Cantor sets with dimension \( d_{f}=\ln 2/\ln 3 \)
for finite generations: \( n=14 \) (square), \( n=13 \) (star),
\( n=12 \) (cross) and \( n=11 \) (plus); (b) Cantor sets with \( d_{f}=\ln 4/\ln 5 \)
for some finite generations: \( n=10 \) (plus), \( n=9 \) (cross),
\( n=8 \) (star) and \( n=7 \) (square); (c) gaskets with \( d_{f}=\ln 3/\ln 2 \)
for finite generations: \( n=11 \) (square), \( n=10 \) (star),
\( n=9 \) (cross) and \( n=8 \) (plus). For all the three cases
\( \alpha =2(d_{f}-d) \) and the dotted lines indicate the best fit
curves of the form \( a(x-b)^{-d} \); where \( d \) is the embedding
dimension of the fractals {[}\( d=1 \) for (a) and (b) and \( d=2 \)
for (c){]}. Insets show the unscaled distributions \( P(m) \) for
overlap \( m \). }{\footnotesize \par}

\subsection{\noindent Overlaps of percolating clusters on square lattice}

\noindent Here we study the overlap distribution of two well-characterised
random fractals; namely the percolating fractals \cite{Stauffer 92}.
Efficient algorithms are available to generate such fractals. It seems,
although many detailed features of the clusters will change with the
changes in (parent) fractals, the subtle features of the overlap distribution
function remains unchanged.

We generate numerically several site percolating clusters at the percolation
threshold (\( p_{c}=0.5927 \) \cite{Stauffer 92}) on square lattices
of linear size \( L \) using Hoshen-Kopelman algorithm \cite{Stauffer 92,Leath 76}.
For the overlap of any two clusters, we count the number of sites
\( M \) which are occupied in both the clusters (see Fig 5). This
gives the overlap size \( m=M/L^{d} \) between the fractals. As the
realisations change (keeping the fractal dimension \( d_{f} \) of
the percolating cluster same) \( m \) varies and we find out its
distribution \( P(m,L) \). We find that the distribution shifts continuously
as \( L \) increases and has a finite width which diminishes but
very slowly with \( L \). This shows, the emerging length scale associated
with \( m \) is no longer a constant, rather it depends on \( L \).
This arises due to the fractal nature of the original clusters, where
the occupation of the sites are no longer random events, but are correlated
\cite{Stauffer 92}. Hence, for a system of size \( L \), the probability
of occupation grows as \( L^{d_{f}} \) for any of the fractals and
as \( [L^{d_{f}}/L^{d}]^{2}=L^{2(d_{f}-d)} \) for the overlap set.
If this is the origin of the \( L \) dependence in \( P(m,L) \),
then the distributions for different \( L \) should follow a scaling
behavior as indicated by eqn. (1).

\resizebox*{7cm}{7cm}{\rotatebox{-90}{\includegraphics{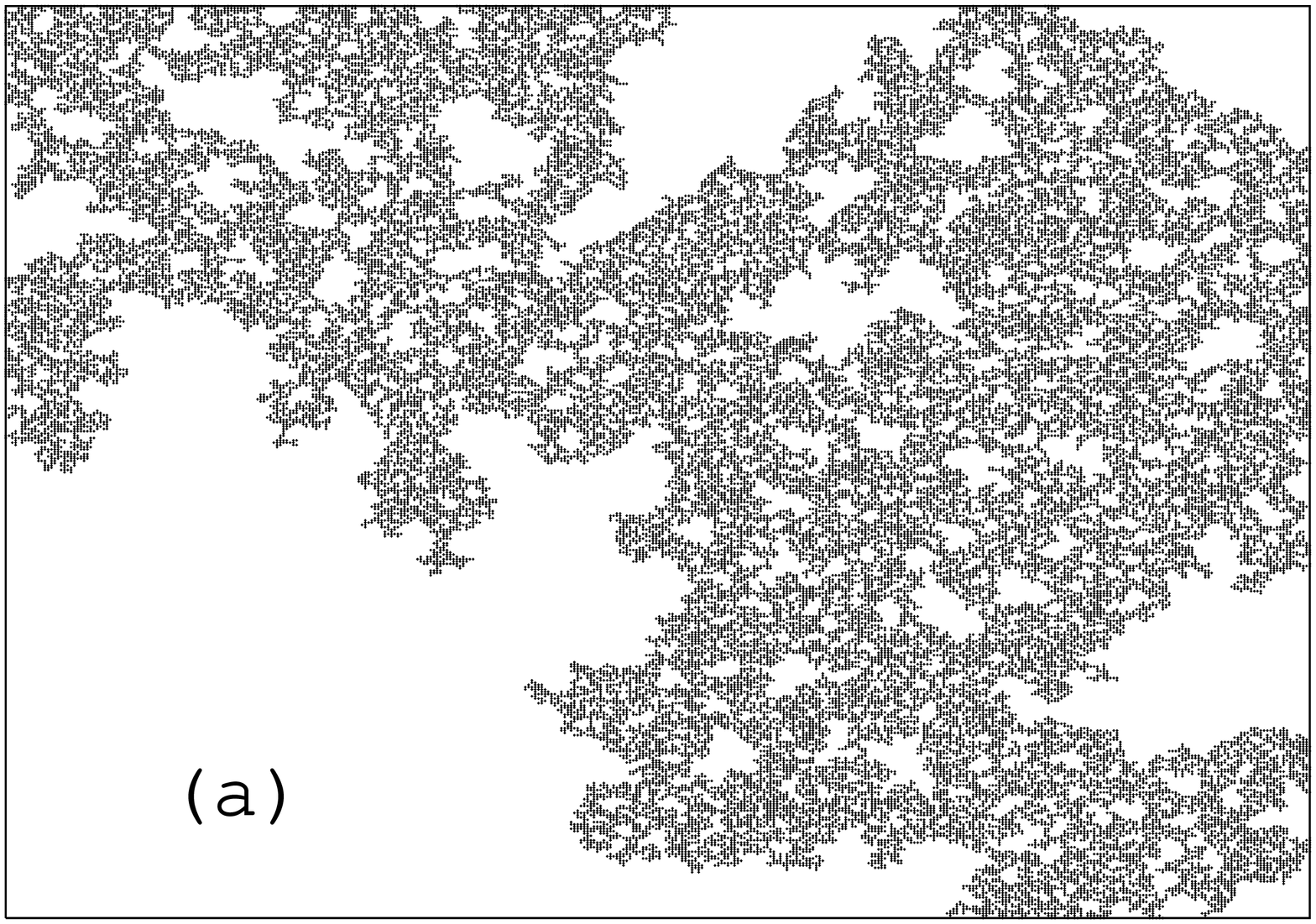}}} 

\resizebox*{7cm}{6cm}{\rotatebox{-90}{\includegraphics{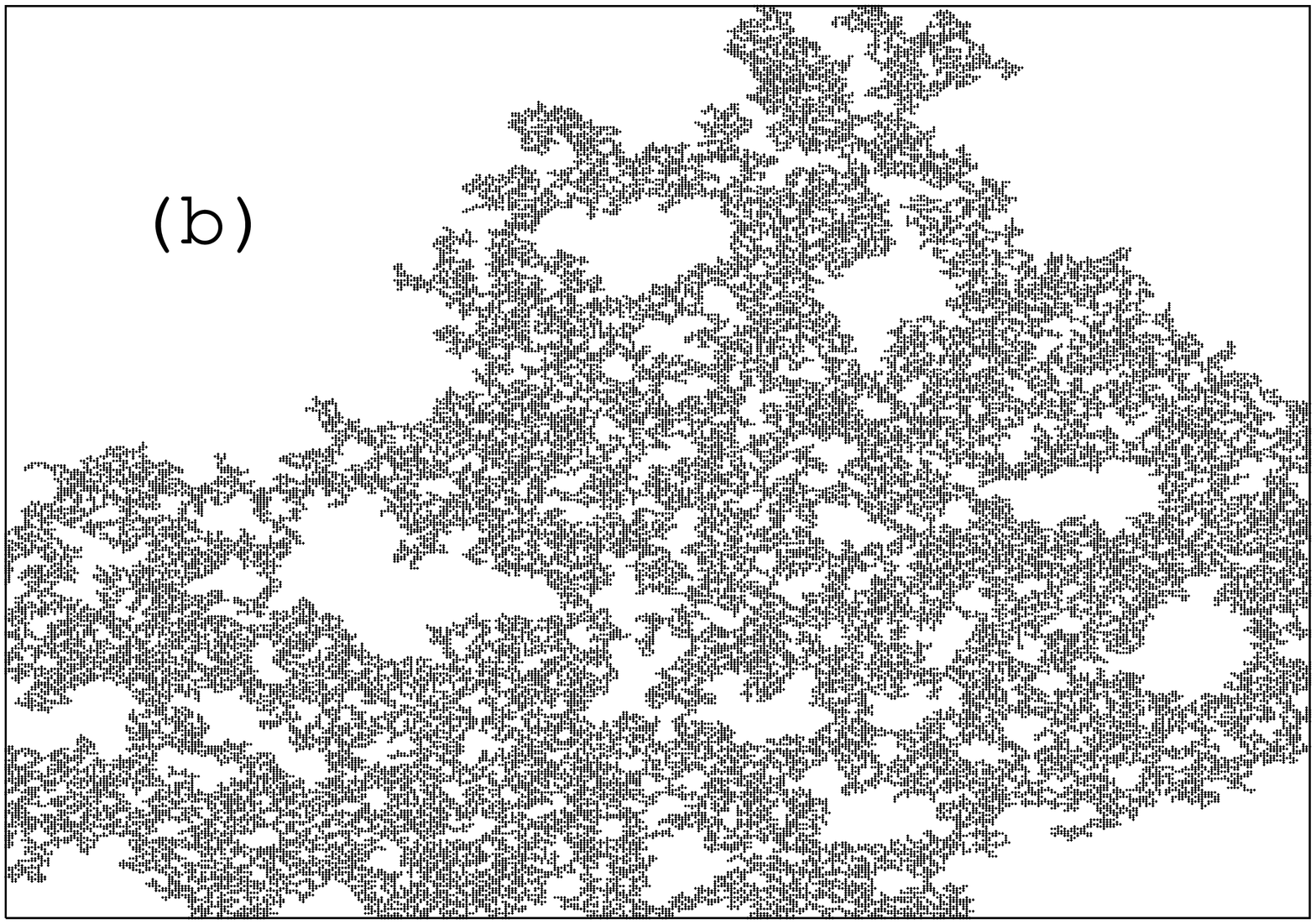}}} 

\vspace{0.3cm}
{\centering \resizebox*{7cm}{5.5cm}{\rotatebox{-90}{\includegraphics{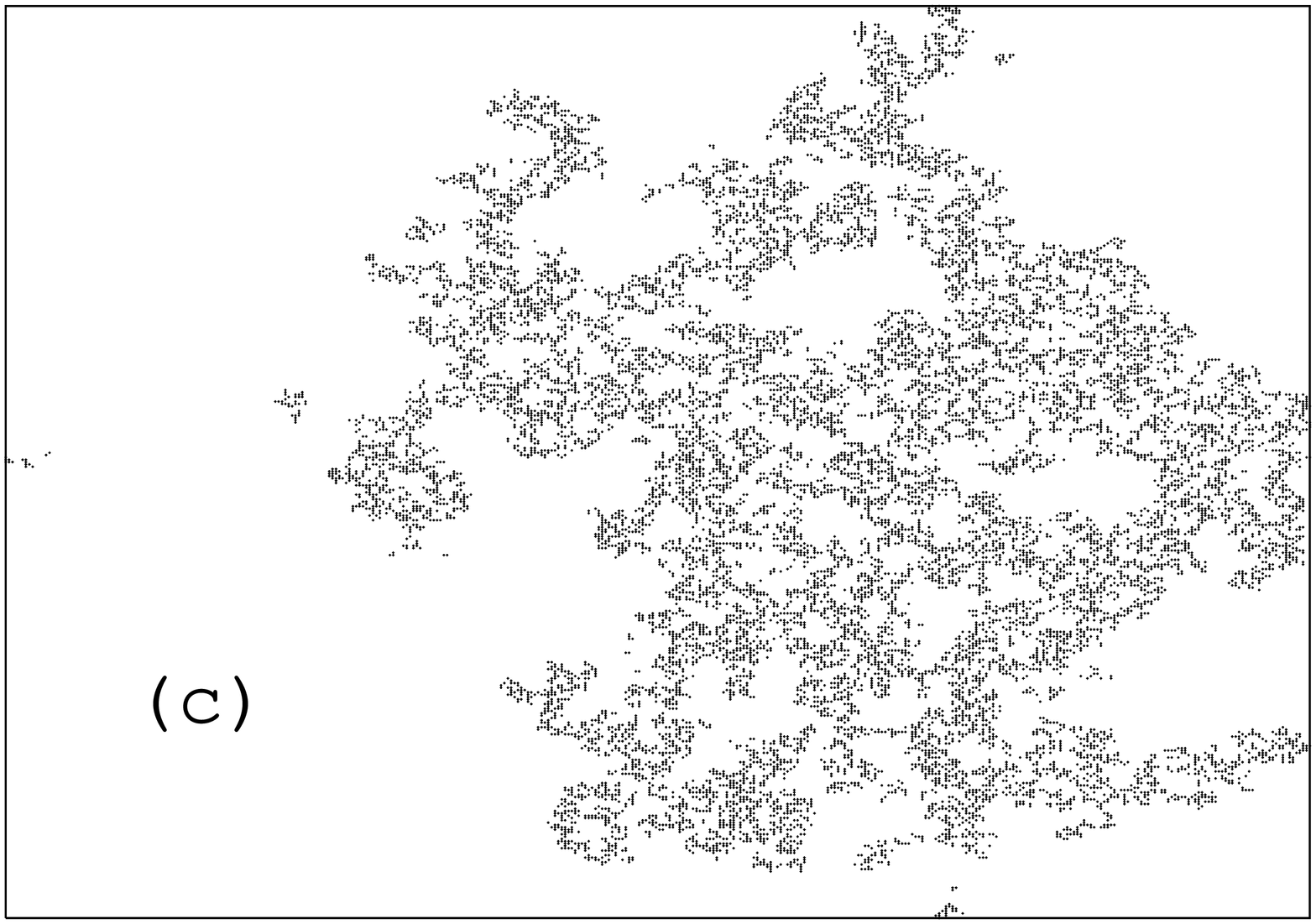}}} \par}
\vspace{0.3cm}

\noindent {\footnotesize Fig 5:} \textbf{\footnotesize }{\footnotesize The
overlap between two percolating clusters; (a) and (b) are two typical
realisations of the same percolating fractal on square lattice (\( d_{f}\simeq 1.89 \))
and (c) their overlap set. Note, the overlap set need not be a connected
one. }{\footnotesize \par}

\vskip.1in

From the overlap between all the pairs of cluster configurations (typically
around 500 for \( L=400 \)), we determine the distribution \( P(m,L) \).
The data are binned to facilitate storage and to make the distribution
smooth (Fig 6 (a)). We have also studied the nature of the distribution
\( P(m) \) for percolation clusters generated with lattice occupation
probability \( p \) above \( p_{c} \) of the square lattice. Here
the distributions become delta functions and the height of the delta
function increases with system size \( L \) (Fig 6 (b)). 

\resizebox*{7cm}{5.5cm}{\rotatebox{-90}{\includegraphics{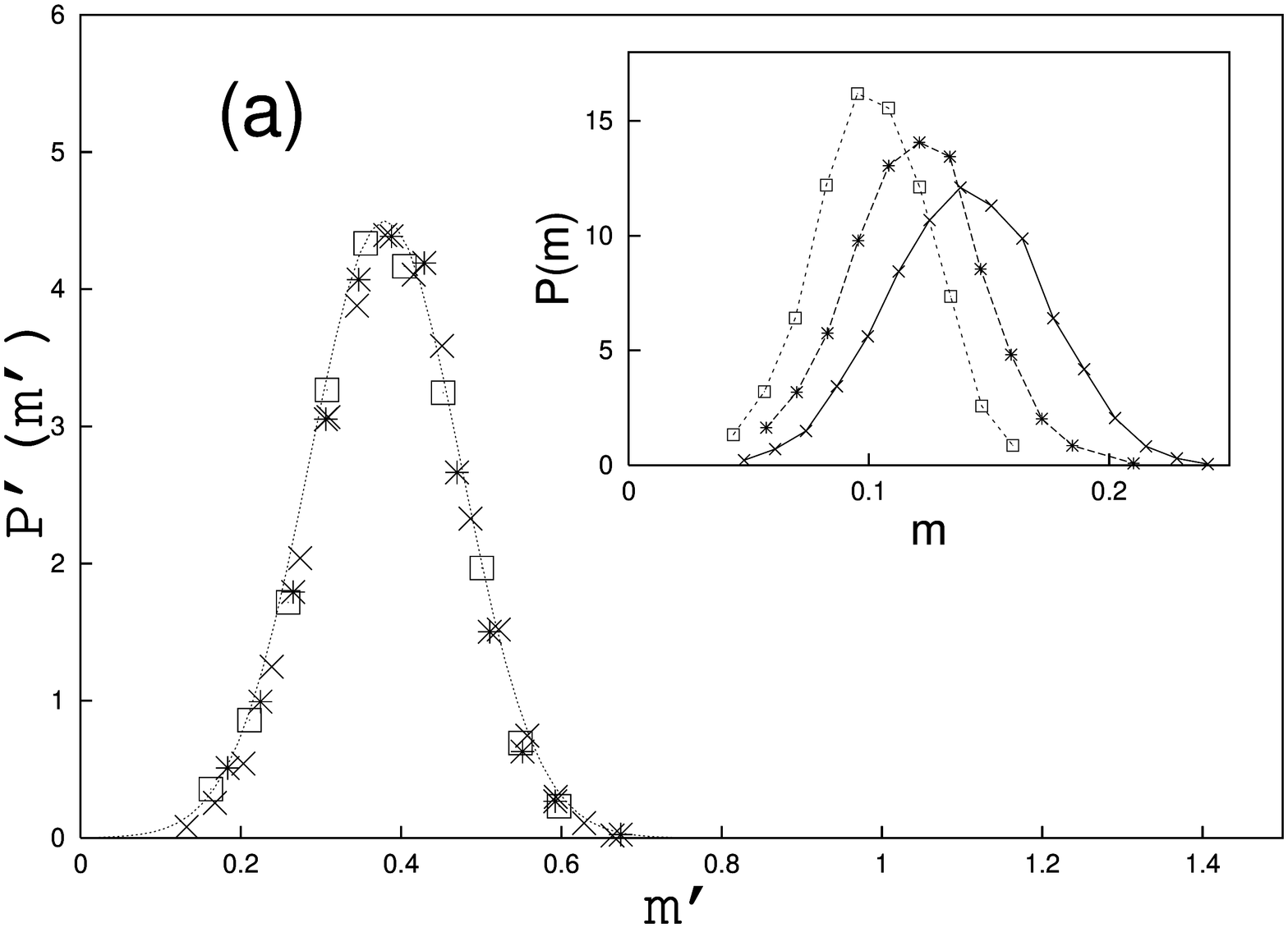}}} 

\resizebox*{8cm}{6cm}{\rotatebox{-90}{\includegraphics{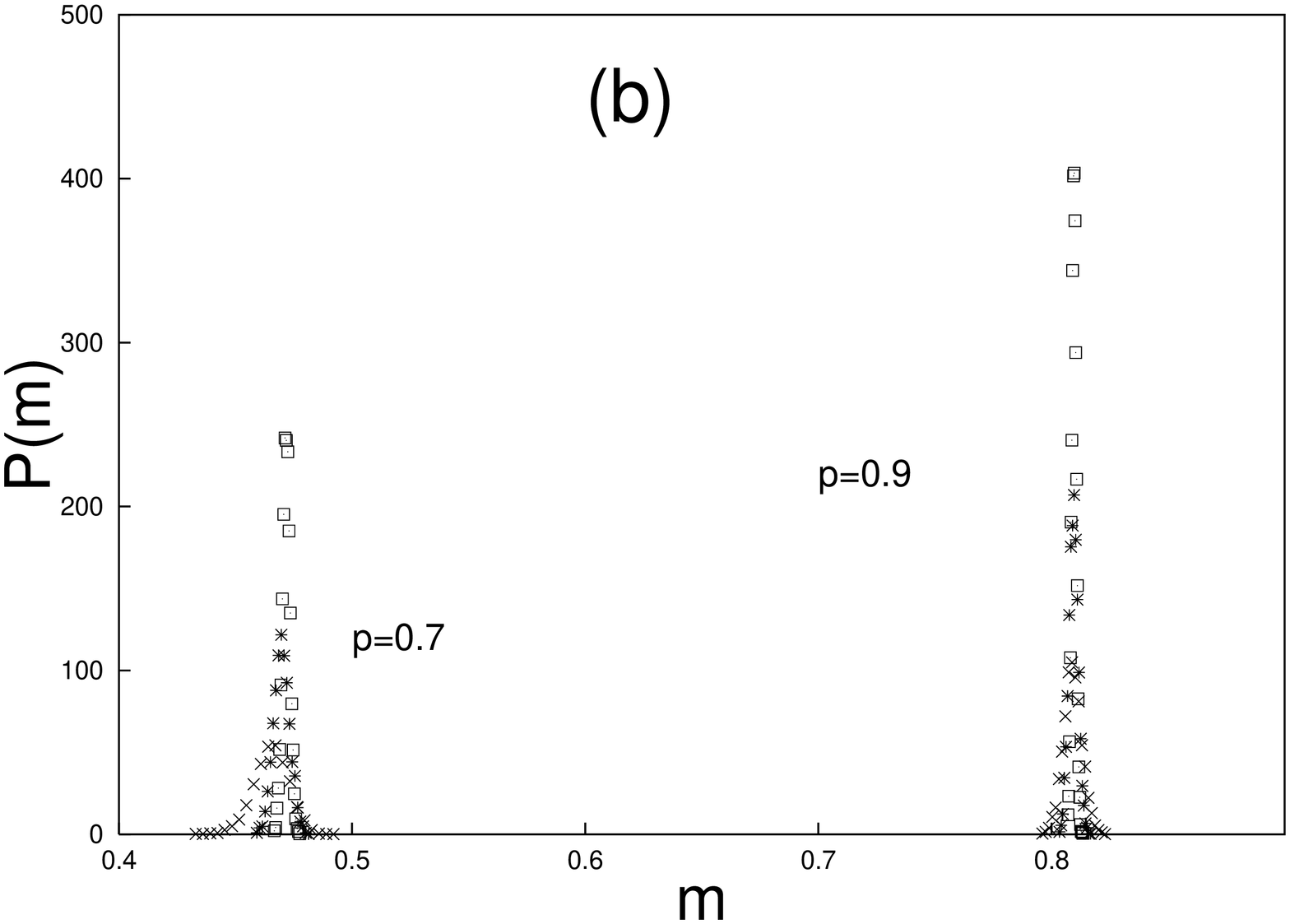}}} 

\vskip.1in

\noindent {\footnotesize Fig 6: (a) The scaled distribution plot of
\( P^{\prime }(m^{\prime })=P(m,L)L^{\alpha } \) against the scaled
overlap \( m^{\prime }=mL^{-\alpha } \) for percolating clusters
grown with probability \( p=p_{c}=0.5927 \) on square lattice (\( d_{f}\simeq 1.89 \))
for finite sizes: \( L=100 \) (cross), \( 200 \) (star) and \( 400 \)
(square). Here \( \alpha =2(d_{f}-d) \). The dotted line indicates
the best fit curve of the form \( a\exp (-(x-b)^{2.0}/c) \); where
\( a \) (\( =4.5 \)) , \( b \) (\( =0.38 \)) and \( c \) (\( =0.018 \))
are constants. Inset shows the unscaled distribution \( P(m) \) against
overlap size \( m \). (b) The distribution \( P(m) \) against overlap
size \( m \) for percolation clusters generated with occupation probability
\( p \) above \( p_{c} \) (\( p=0.7 \) and \( p=0.9 \)) on square
lattice {[}for finite sizes: \( L=100 \) (cross), \( 200 \) (star)
and \( 400 \) (square){]}. Clearly the distributions \( P(m) \)
are delta functions and the height of the delta function increases
with the system size \( L \). }{\footnotesize \par}

\vskip.2in

\section{Conclusion}

\noindent In the context of studying the earthquake magnitude distribution,
we have studied here a model where the earthquake fault and the moving
tectonic plates are assumed to be self-similar fractals. Essentially,
we study the contact area distribution of two nominally identical
fractals. Our study on different sets of fractal surface overlap shows
that the contact area distributions are not always power laws. Although
for Cantor sets and gaskets having different fractal dimensions (both
for regular and random cases), the contact area distributions have
got power laws, for percolating clusters (at percolation threshold)
it takes a robust Gaussian form which we can't explain even qualitatively.
In particular, we find that in all the cases the distributions \( P(m,L) \)
show an universal finite size (\( L \)) scaling behavior \( P^{\prime }(m^{\prime })=L^{\alpha }P(m,L) \);
\( m^{\prime }=mL^{-\alpha } \), where \( \alpha =2(d_{f}-d) \).
The \( P(m) \), and consequently the scaled distribution \( P^{\prime }(m^{\prime }) \),
have got a power law (2) decay with \( m \) (with decay exponent
equal to \( d \)) for both regular and random Cantor sets and also
for gaskets. For percolation clusters, \( P(m) \) (and hence \( P^{\prime }(m^{\prime }) \))
have a Gaussian variation with \( m \) (Fig. 6a). It may be mentioned
that, since in many cases the bulk thermodynamics may be mapped to
some reduced dimensional interface problems and since their overlaps
corresponds to physical quantities (e.g. , the replica overlap in
the physics of glasses \cite{Young 99}), such fractal overlap distributions
are of quite general interest.

\vskip -.10 in

\textbf{Acknowledgement}: We are grateful to Prof. D. Stauffer for
his useful comments and for a critical reading of the manuscript.
We are also thankful to A. Hansen and J. Kertez for their comments.

\end{document}